\documentclass[aps,prl,twocolumn,showpacs,superscriptaddress,superscriptaddress]{revtex4-1}
\usepackage{graphicx}  
\usepackage{bm}        
\usepackage{amssymb}   
\usepackage{amsmath}
\usepackage{bm}
\usepackage{bbm}
\usepackage{tikz}
\usepackage{pgfplots}
\usepackage{caption}

\usepackage{subcaption}
\usepackage{todonotes}

\newlength\figurewidth
\newlength\figureheight

\newlength\mywidth
\newcommand\A{\ensuremath{\bm{A}}}
\newcommand\x{\ensuremath{\bm{x}}}
\newcommand\m{\ensuremath{\bm{m}}}

\newcommand\n{\ensuremath{\mathbf{n}}}
\newcommand\R{\ensuremath{\mathbb{R}}}

\newcommand\bn{\ensuremath{\bm{\nabla}}}

\renewcommand\i{\ensuremath{\textup{i}}}

\newcommand\dfn{\ensuremath{\mathrel{\mathop:}=}}

\newcommand\tp{\ensuremath{\mathrm{T}}}

\newcommand\J{\ensuremath{\mathcal{J}}}
\newcommand\K{\ensuremath{\mathcal{K}}}
\newcommand\PP{\ensuremath{\mathcal{P}}}
\newcommand\conj[1]{\ensuremath{\overline{#1}}}

\begin{document}
\title{Efficient determination of the energy~landscape of~nonlinear~Schr\"odinger-type~equations}

\author{Nico Schl\"omer}
\affiliation{Departement Wiskunde en Informatica, Universiteit
Antwerpen, 2020 Antwerpen, Belgium}
\author{Daniele Avitabile}
\affiliation{School of Mathematical Sciences, University of Nottingham, Nottingham, UK}
\author{Milorad V. Milo\v{s}evi\'c}
\author{Bart Partoens}
\affiliation{Departement Fysica, Universiteit Antwerpen,
Groenenborgerlaan 171, 2020 Antwerpen, Belgium}
\author{Wim Vanroose}
\affiliation{Departement Wiskunde en Informatica, Universiteit
Antwerpen, 2020 Antwerpen, Belgium}

\begin{abstract}
We describe a systematic approach for the \emph{efficient numerical
solution} of nonlinear Schr\"odinger-type partial differential
equations of the form $(\K +V + g|\psi|^2)\psi=0$, with an
energy operator $\K$, a scalar potential $V$, and a scalar parameter $g$.
Instrumental to the approach
are developments in numerical linear and nonlinear algebra,
specifically numerical parameter continuation. We demonstrate how a
continuous sequence of solutions can be obtained regardless of their
stability, so that finally the \emph{spectrum of stable and
unstable solutions} in the specified parameter range is fully
revealed. The method is demonstrated for the GL
equation in a three-dimensional superconducting domain with an inhomogeneous
magnetic field, a numerically demanding problem known to have an
involved solution landscape.
\end{abstract}

 \maketitle

Nonlinear Schr\"odinger equations and their variations are used to
model a wide variety of physical systems \cite{tsal,rcf},
with applications spanning superconductivity \cite{glt}, quantum condensates
\cite{gpe1,gpe2}, nonlinear acoustics \cite{som1979coupled},
nonlinear optics \cite{gedalin1997optical}, and hydrodynamics
\cite{nore1993numerical}. Typically, the physical models described
by these equations contain a set of parameters specifying,
e.g., the sample geometry, external fields, or boundary conditions.
To understand the physical properties of the system, it is interesting
to explore the energy landscape of the steady state solutions as a
function of one or more of these parameters. In general, slight perturbations in one
of the control parameters can induce changes in the stability properties of the
states, causing abrupt transitions in the energy landscape \cite{PhysRevB.65.104515}.
Although the
existence of steady states can be proven in certain cases,
analytic solutions are hard or impossible to obtain in realistic systems. Numerical
methods are hence of particular importance for
understanding the physics of the systems modeled by nonlinear
Schr\"odinger equations. The challenge is to develop efficient
computational tools to explore the full energy landscape, including
minima and saddle-points, of three- and higher-dimensional systems.

In general, the nonlinear Schr\"odinger equations describing the evolution of a
quantum-dynamical system represented by a complex-valued order parameter $\psi$
are written as
\begin{equation}\label{eq:tschroed}
\begin{cases}
\i\dfrac{\mathrm{d}\psi}{\mathrm{d}t} =
\mathcal{S}(\psi)\dfn
(\K + V + g|\psi|^2)\psi,\\[1.5ex]
\psi(0) = \psi_0,
\end{cases}
\end{equation}
with a linear, positive-semidefinite (energy) operator~$\K$, an
external potential~$V$, and the coupling parameter~$g$.
The term $|\psi|^2$ usually describes a probability density of the model entity,
e.g., the locality of quantum particles. Examples include
the Gross--Pitaevskii equation
where $\K=-\Delta$ is the negative Laplacian,
and the Ginzburg--Landau (GL) equation, where $\K =\left(-\i\bn -
\A\right)^2$ is the covariant Laplacian with a given vector potential $\A$.

To understand the long-term dynamics, it is essential to compute steady states of the
system, i.e., solutions to
\begin{equation}\label{eq:schroed}
0 = \mathcal{S}(\psi).
\end{equation}
Solving~\eqref{eq:schroed} in realistic three-dimensional domains is a difficult numerical
task: the number of unknowns quickly becomes very large and standard
numerical methods become impractical. In addition, the energy landscape of the
solutions becomes very complicated and its systematic exploration is prohibitive with
current numerical techniques.

This letter describes a numerical approach for the efficient computation of the
steady-state landscape as a function of the control parameters. Its central component
is a Newton--Krylov algorithm
\cite{kelley1995iterative, knoll2004jacobian} to solve the nonlinear
problem~\eqref{eq:schroed}, combined with
numerical parameter continuation
\cite{Krauskopf2007} to explore the solution landscape.
Although numerical continuation and Newton--Krylov solvers are well-known methods
for large-scale systems, they cannot be applied straightforwardly
to~\eqref{eq:schroed}. We
will show  that, by exploiting the
properties of the linearization
of the operator $\mathcal{S}(\cdot)$ and devising a specially tailored
preconditioner, it is possible to considerably accelerate the convergence of the
linear iterations. This opens up the possibility to compute steady states and
to systematically explore the energy landscape in three-dimensional problems.
The new method scales
optimally with the number of unknowns in the system and is fully
parallelizable.
In particular, we illustrate the power of the
method  by studying three-dimensional vortex nucleation in an extreme-type--II
superconductor in an inhomogeneous magnetic field,
described by the GL equation.

Numerical simulations within the GL model are an essential tool for the analysis of
superconducting phenomena. In this area, vortex matter has been at the forefront of
research in the past two decades. Emphasis has been put on the computation of vortex
states with imposed confinement (i.e., the sample shape) and on their dependence upon
critical parameters of the superconducting sample.
Of particular relevance are unstable states of the system, often called
\emph{saddle points}. These solutions shape the energy landscape as they
constitute the connections between families of stable states, thereby providing a
unique insight into the dynamic transitions and vortex rearrangements that have been
observed experimentally.
Notably, with the framework proposed in this letter we can compute both stable and unstable
states, which are not accessible with traditional numerical methods.
Owing to numerical difficulties, saddle points have been calculated only for
radially-symmetric samples such as disks, using a limited-expansion method
\cite{schw1999}. Recently, the full energy landscape (including saddle points) was
systematically explored in two-dimensional square samples, using numerical
continuation techniques~\cite{SAV:2012:NBS}: in particular, it was possible to build
an atlas of the instabilities occurring in the sample, providing a complete
classification of the symmetries of observable stable states. In this letter, we
address the much more challenging case of three-dimensional samples of arbitrary
shape.

\paragraph{Existing methods} The literature on numerical
methods for the time-dependent equation \eqref{eq:tschroed} is
rather extensive and mostly concerned with time-stepping schemes
\cite{Taha1984203,sanz1984methods,Chang1999397}. For example,
references \cite{PhysRevA.51.4704,bao2003numerical} leveraged specific
properties of certain numerical procedures and settings for dealing
with the Gross--Pitaevskii equation.
Stationary-states are typically found by applying a
(pseudo-)time-stepping scheme until a stationary state is reached
\cite{schwPRL,PhysRevLett.79.4653,geurts}. There are, however,
several disadvantages with this approach.
Firstly, iterations converge only for strictly stable
states, therefore unstable or saddle point states can not be computed.
Secondly, stable solutions may have extremely long (sometimes oscillatory) transients,
therefore convergence for three-dimensional domains may be prohibitively slow.

\paragraph{Newton's method}
%
A better approach is to use Newton iterations directly on~\eqref{eq:schroed},
starting from a suitable initial guess: Newton's method converges
superlinearly in a neighborhood of the solution, irrespectively of the
stability properties of the equilibrium. Once a
steady state $\psi$ is found, stability is determined by computing the spectrum of
the operator 
obtained by linearizing $\mathcal{S}$ around $\psi$. For the Schr\"odinger equations,
this linear operator is defined via the action
%
\begin{equation}\label{eq:schroed J}
\J(\psi)\phi = (\K + V + 2g|\psi|^2) \phi + g \psi^2 \overline{\phi},
\end{equation}
where $\overline{\phi}$ denotes complex conjugation.

A sequence of approximations $\psi^{(k)}$ to the steady state is computed with
Newton's method $\psi^{(k+1)} = \psi^{(k)} + \delta \psi^{(k)}$, where the update $\delta
\psi^{(k)}$ satisfies
\begin{equation}\label{eq:newton}
\J(\psi^{(k)}) \delta \psi^{(k)} = -\mathcal{S}(\psi^{(k)}).
\end{equation}
Therefore, the solution of a large linear system is required at each Newton step $k$, which
is the most significant difficulty when applying Newton's method to
nonlinear Schr\"odinger equations.


\paragraph{Solving the Jacobian system}

Linear systems such as~\eqref{eq:newton} can be solved using
Krylov iterative methods and have been widely used in the past decades
\cite{saad2003iterative,greenbaum1997iterative}.
A property of Krylov subspace methods is
that no explicit (matrix) representation
of the operator is needed, but only its
application to vectors (cf.~(\ref{eq:schroed J})).
The convergence of those methods is highly dependent
on the spectrum of the involved linear operator.
Principal optimizations can be employed
if all eigenvalues of the respective linear operator
are real-valued.
This is the case if the linear operator is represented by a Hermitian matrix or, in
general, if $\J$ is self-adjoint
with respect to a given inner product.
The linear operator~(\ref{eq:schroed J}) associated with the nonlinear Schr\"odinger
equations is self-adjoint with respect to the inner product
\begin{equation}\label{eq:scalar product}
\langle \phi, \psi\rangle \dfn \Re \left( \int_\Omega \conj{\phi} \psi \right).
\end{equation}
This suggests the use of MINRES \cite{greenbaum1997iterative}, a
Krylov subspace method suitable for indefinite self-adjoint
problems. However, one characteristic of Krylov methods is that a larger
number of unknowns increases the number of iterations that are needed to achieve convergence.
In addition, the computational cost of a single evaluation of the linear
action also grows with the number of unknowns. Therefore, high-resolution
discretizations of three-dimensional systems would require a prohibitive
computational effort.
Indeed, decreasing the number of Krylov iterations is the subject of extensive
research efforts in this area.

A popular approach is to use a \emph{preconditioner} for the linear problem.
The main idea is that, instead of solving  the discretized version $J x = b$ of
\eqref{eq:newton}, one can solve an equivalent, numerically more favorable problem
$P^{-1}J x = P^{-1}b$ with a linear, invertible preconditioning operator $P$.
%
If $P$ is appropriately chosen, Krylov methods applied
to the new system converge much faster. In the case of the linearization of
nonlinear Schr\"odinger operators (\ref{eq:schroed J}), the energy
operator $\K$ is of particular interest, as it typically dominates the spectral
behavior of $\J(\psi)$. More precisely, we
define the symmetric preconditioning operator
\begin{equation}\label{eq:prec}
\PP(\psi) \dfn \K + 2\max\{g,\varepsilon\}|\psi|^2,
\end{equation}
with $0<\varepsilon\ll 1$ \cite{SV:2012:OLS}. We note that $\PP(\psi)$ is positive-definite except
for the uninteresting case of $\psi\equiv 0$. This, most importantly,
makes the inversion of the discretized $\PP(\psi)$, $P^{-1}(\psi)$, computationally cheap
since its positive-definiteness makes it a
suitable target for geometric or algebraic multigrid (AMG) solvers
that yield optimal convergence \cite{trottenberg2001multigrid}. As
will be shown, even inexact inversions of~(\ref{eq:prec}) used
as preconditioners for~(\ref{eq:schroed J}) make the Krylov
convergence \emph{independent of the number of unknowns.}

\paragraph{Numerical parameter continuation}

The efficient linear solver outlined above is an essential building block for the
exploration of the energy landscape, which is performed via numerical parameter
continuation, a well-established technique for numerical bifurcation analysis of
dynamical systems \cite{keller1987lectures}.
Let $F(\psi;p)=0$ be a nonlinear system dependent upon a scalar parameter $p \in \R$ and
let $\psi^*_0$ be a solution for a given parameter value $p^*_0$. Under generic regularity
conditions for $F$, it is possible to construct a one-parameter family of solutions
$\psi$, parametrized by $p$, in the neighborhood of $(\psi^*_0,p^*_0)$
\cite{Krauskopf2007}. First, a prediction step $(\psi_1,p_1)$ is taken in the tangent
direction to the one-parameter family, then a correction is done using
Newton's method. This leads to a new solution $(\psi^*_1,p^*_1)$. The set of
points $(\psi^*_k,p^*_k)$ form a smooth solution curve. Once again, we remark
that the method is oblivious to the stability properties of the solutions.

\paragraph{Application to the Ginzburg--Landau problem}

We illustrate the power of this method on a numerically challenging problem:
the computation of vortex structures in a \emph{three-dimensional} superconducting
domain with a magnetic core, which establishes an \emph{internal} and
\emph{inhomogeneous} magnetic field.

Given a bounded superconducting domain
$\Omega$, the GL equations
\begin{equation}\label{eq:GL compact}
\begin{split}
0 &=  \left(-\i\bn - \A\right)^2 \psi - \psi \left(1 - |\psi|^2\right) \quad \text{in } \Omega,\\
0 &=  \n \cdot ( -\i\bn - \A) \psi \quad \text{on the surface}~\partial\Omega.
\end{split}
\end{equation}
describe stationary states of an extreme-type-II
superconductor subject to a magnetic field associated with the vector
potential $\A$ \cite{schwPRL,DGP:1992:AAG}.
The equations are presented in dimensionless form: distances are scaled by the
superconducting coherence length $\xi$, the order parameter $\psi$
by its value in the absence of applied magnetic field, and the vector
potential by $\xi H_{c2}$, where $H_{c2}$ denotes the upper
critical magnetic field of a bulk material.
Since the kinetic
energy operator $(-\i\bn-\A)^2$ is Hermitian and positive
semi-definite, equation~\eqref{eq:GL compact} is of the form
\eqref{eq:schroed} (with $V\equiv-1$, $g=1$)
and can be solved with our numerical method.

\begin{figure}
\setlength\figurewidth{0.6\mywidth}
\includegraphics[width=\figurewidth]{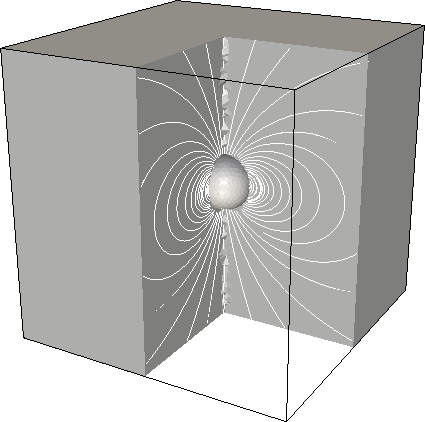}
\caption{Cube with a spherical cavity, representing a
  superconducting sample hosting a magnetic dipole (clipped
  display). The field lines of the magnetic field $\bn\times\A_\text{d}$
  of the dipole are shown in white.}
\label{fig:domain}
\end{figure}

We choose $\Omega$ to be a cube with side length $10$
and a spherical cavity of radius~$1$ containing a magnetic dipole
with magnetic moment $\m=(0,0,1)^\tp$
(see figure~\ref{fig:domain}). The
associated magnetic vector potential of such a dipole is
$\A_\text{d}(\x) \dfn \|\x\|^{-3} (\m\times\x)$.
This example is of great relevance to the field, since
the expected loops appear in
several physical systems \cite{ScVort}. Their
nucleation, growth, motion, and recombination harbors
a vast variety of novel physics.
We perform numerical experiments using a finite-volume
(tetrahedral) discretization where the complex-valued order
parameter $\psi$ is approximated in the grid nodes \cite{SV:2012:OLS}.

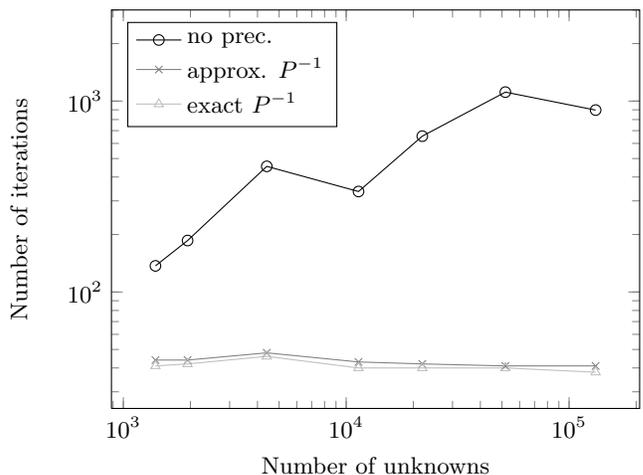
\begin{figure}
\centering
\setlength\figurewidth{1.2\mywidth}
\setlength\figureheight{0.8\figurewidth}
\begin{tikzpicture}

\begin{loglogaxis}[
xlabel={Number of unknowns},
ylabel={Number of iterations},
ymin=0, ymax=3000,
width=\figurewidth,
height=\figureheight,
legend style={legend pos=north west, legend cell align=left}
]
\addplot [black, mark=o]
coordinates {
(1394,137)
(1943,186)
(4403,455)
(11359,336)
(21952,655)
(51782,1114)
(131550 ,897)
};
\label{addplot:unprec}
\addlegendentry{no prec.}

\addplot [gray, mark=x]
coordinates {
(1394,44)
(1943,44)
(4403,48)
(11359,43)
(21952,42)
(51782,41)
(131550 ,41)
};
\addlegendentry{approx. $P^{-1}$}
\label{addplot:prec1}

\addplot [lightgray, mark=triangle]
coordinates {
(1394,41)
(1943,42)
(4403,46)
(11359,40)
(21952,40)
(51782,40)
(131550 ,38)
};
\addlegendentry{exact $P^{-1}$}
\label{addplot:prec2}

\end{loglogaxis}

\end{tikzpicture}
\caption{The number of MINRES iterations necessary for converging the
initial guess $\varphi_0\equiv 0$ for the system
$\J(\psi)\varphi=b$, $\psi\equiv 1$, $b\equiv 1$ to a relative
residual of $10^{-10}$ for different preconditioners.
The figure shows that the number of iterations increases
without a preconditioner (\ref{addplot:unprec}),
and remains constant for preconditioners with exact (\ref{addplot:prec2})
as well as approximate inversion (\ref{addplot:prec1}).}
\label{fig:prec}
\end{figure}

The first important result is shown in figure~\ref{fig:prec}, and concerns the
efficiency of solving the Jacobian systems with preconditioners based on the
discretization $P(\psi)$ of \eqref{eq:prec}. More specifically, we use two
preconditioners, the first one being the exact inverse (up to machine precision) of
$P(\psi)$ and the second one being an approximate inverse of $P(\psi)$ obtained with
just a single AMG step. Preconditioned linear systems are solved for increasing
number of unknowns and they are compared to the case without preconditioners.
A remarkable result is that, in both preconditioned cases, the number of iterations
\emph{does not increase} with the number of unknowns in the system. This indicates optimal
scalability of the solver, which is extremely advantageous compared to the case
without preconditioner.

\begin{figure}
\centering
\setlength\figurewidth{0.55\mywidth}
\input{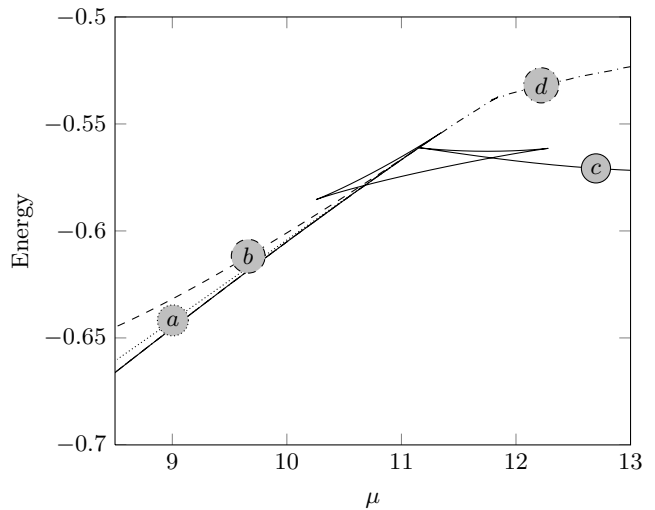}
\caption{Numerical continuation in the parameter $\mu$
for the cube with cavity as depicted in figure~\ref{fig:domain}.
The Gibbs energy diagram
elucidates four solution branches spawning from one base branch.
The branch (\ref{addplot:4}) corresponds to the superconducting Meissner
state and contains $(\psi_0;0)$ when continued to smaller $\mu$. Representative
solutions on each branch are shown in figure~\ref{fig:illust}.}
  \label{fig:branches}
\end{figure}

\begin{figure}
  \subcaptionbox{\label{subfig:sol1}}{\includegraphics[width=0.55\mywidth]{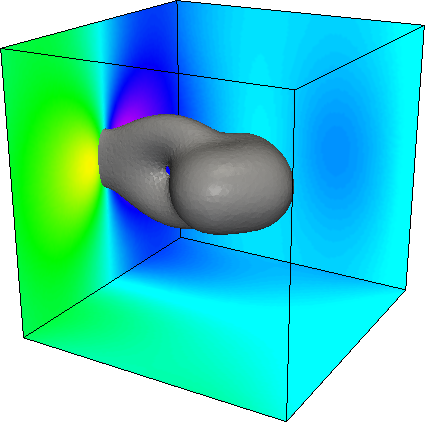}}\hfill
  \subcaptionbox{\label{subfig:sol2}}{\includegraphics[width=0.55\mywidth]{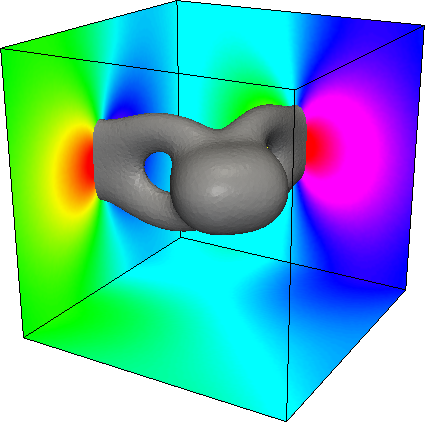}}\\
  \subcaptionbox{\label{subfig:sol3}}{\includegraphics[width=0.55\mywidth]{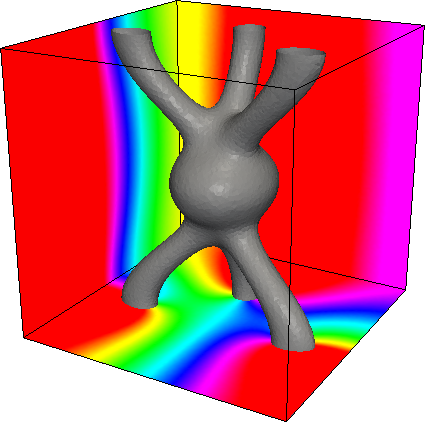}}\hfill
  \subcaptionbox{\label{subfig:sol4}}{\includegraphics[width=0.55\mywidth]{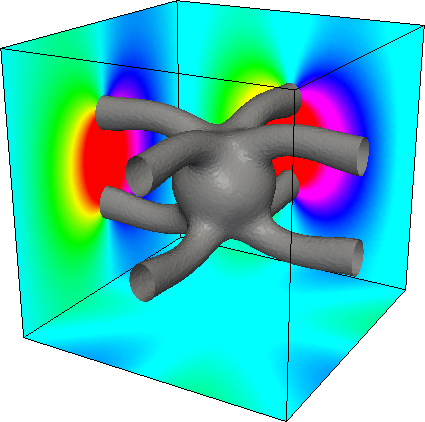}}
  \caption{ Exemplary states from figure~\ref{fig:branches}. Shown are the isosurfaces of
    $|\psi|^2=0.1$ and the phase of the superconducting order parameter on
    three of the faces of the domain. The field $|\psi|^2$ is strongly depleted
    near the magnetic dipole, at the center, where vortex loops emerge and stretch
    out to the boundary. For details, see animated data in the supplementary
    material.}\label{fig:illust}
\end{figure}

Numerical parameter continuation is then applied to a discretization
with $0.4\times10^6$ grid points.
The magnetic vector potential
$\A_\text{d}$ (and thus the corresponding magnetic field $\bn\times\A_\text{d}$) is
scaled with the dimensionless magnetic moment
$\mu=m/(H_{c2}\xi^3)$ which is taken as control parameter. For $\mu=0$, the homogeneous
state $\psi\equiv 1$ is clearly a solution (independently of the
domain) and can be used to start off the parameter continuation. Alternatively,
the computation can be started from a solution obtained otherwise (e.g., via
time stepping).
As
shown in figure~\ref{fig:branches}, our method automatically generates the energy
landscape, parametrized by $\mu$, revealing the existence of several
branches with different energy (for a definition of Gibbs free energy
see, e.g., \cite{schwPRL,geurts}). Initially, at low $\mu$, the
superconducting order parameter is strongly suppressed only around
the embedded dipole. For increasing $\mu$, vortex loops emerge from
this area, connecting the poles of the dipole. Initially, exactly 4 such loops
are present in the stable solution
which enjoys the fourfold symmetry of the problem (similar to figure~\ref{subfig:sol4}).
A bifurcation occurs at $\mu\approx10.2$, suppressing three loops towards the dipole
and generating a branch of states with just one loop
%
(see supplementary material for the
animation, figure \ref{subfig:sol1} and branch $a$ in figure~\ref{fig:branches}).
Similar behavior is found for the state with two loops, with a bifurcation point at
$\mu\approx10.9$ (see the state in figure~\ref{subfig:sol2} and its
corresponding energy branch). However, none of these states reaches the ground
state of the system: it is actually the three-loops state that prevails
at $\mu\approx11.4$, formed via shrinking one loop along the saddle point and growing
three loops until they hit the cube sides. When $\mu$ is further
increased, these loops follow another saddle point, where they hit the
top and bottom surfaces, and then stabilize as three vortices piercing the sample
top to bottom (figure~\ref{subfig:sol3}). As we continue the computation for higher
values of $\mu$, more bifurcations occur on each energy branch, and new solutions
branches emerge. Since this calculation is only intended to illustrate our approach,
we do not include details about such branches.
The entire solution curves can be obtained from \cite{schloemer:figshare}.


\paragraph{Conclusion and outlook}
In this letter, we developed a computational tool that allows
the efficient exploration of the steady-state landscape of nonlinear
Schr\"odinger-type equations on a high-resolution three-dimensional grid.
It uses a
preconditioned Newton--Krylov solver in combination with a numerical
parameter continuation method. The main advantage is that the
computational cost increases only linearly with the number of
grid points in the calculation. This is due to the fact that the number of
required iterations is independent of the dimension of the
solution space, i.e., the number of unknowns.
As a result, it is now possible to efficiently study
three-dimensional physical systems described by nonlinear
Schr\"odinger equations even on low-end workstations. Note that our
solver 
is entirely built of existing open-source components. In particular we used the
high-performance continuation solver implemented in LOCA \cite{1089021}.

By using Newton's method, our approach gives insight into saddle
point states which are essential for understanding the dynamics of
the GL system. At the same time, the finite-volume discretization
can be used on samples with arbitrary shape. We have employed our solver for a
challenging numerical problem, computing several stable and saddle
point vortex(-loop) configurations in a three-dimensional
superconducting cube encapsulating a magnetic dipole \cite{mauro}.

In conclusion, our method gives access to the dynamics of systems that
were to date perceived to be too complex to be tackled with numerical continuation.
It is inherently applicable to all physical systems modeled by nonlinear
Schr\"odinger equations~\eqref{eq:tschroed}, including the Gross--Pitaevskii equations
for Bose--Einstein condensates, nonlinear optics, plasma physics, deep water waves,
and even seemingly distant subjects such as cosmology and particle
physics. Just as demonstrated for vortices in superconductors, numerical continuation
methods can be used for systematic studies of other
topological solitons (even three-dimensional knotted ones \cite{knot1,knot2}),
solitary waves and breathers, which are common to various equations of nonlinear
Schr\"odinger type (see, e.g., \cite{nonlin}).

\paragraph{Acknowledgment}
This research was supported by Flemish Science Foundation
(FWO~Vlaanderen) through project G.0174.08N.

\bibliography{gl}

\begin{thebibliography}{35}%
\makeatletter
\providecommand \@ifxundefined [1]{%
 \@ifx{#1\undefined}
}%
\providecommand \@ifnum [1]{%
 \ifnum #1\expandafter \@firstoftwo
 \else \expandafter \@secondoftwo
 \fi
}%
\providecommand \@ifx [1]{%
 \ifx #1\expandafter \@firstoftwo
 \else \expandafter \@secondoftwo
 \fi
}%
\providecommand \natexlab [1]{#1}%
\providecommand \enquote  [1]{``#1''}%
\providecommand \bibnamefont  [1]{#1}%
\providecommand \bibfnamefont [1]{#1}%
\providecommand \citenamefont [1]{#1}%
\providecommand \href@noop [0]{\@secondoftwo}%
\providecommand \href [0]{\begingroup \@sanitize@url \@href}%
\providecommand \@href[1]{\@@startlink{#1}\@@href}%
\providecommand \@@href[1]{\endgroup#1\@@endlink}%
\providecommand \@sanitize@url [0]{\catcode `\\12\catcode `\$12\catcode
  `\&12\catcode `\#12\catcode `\^12\catcode `\_12\catcode `\%12\relax}%
\providecommand \@@startlink[1]{}%
\providecommand \@@endlink[0]{}%
\providecommand \url  [0]{\begingroup\@sanitize@url \@url }%
\providecommand \@url [1]{\endgroup\@href {#1}{\urlprefix }}%
\providecommand \urlprefix  [0]{URL }%
\providecommand \Eprint [0]{\href }%
\providecommand \doibase [0]{http://dx.doi.org/}%
\providecommand \selectlanguage [0]{\@gobble}%
\providecommand \bibinfo  [0]{\@secondoftwo}%
\providecommand \bibfield  [0]{\@secondoftwo}%
\providecommand \translation [1]{[#1]}%
\providecommand \BibitemOpen [0]{}%
\providecommand \bibitemStop [0]{}%
\providecommand \bibitemNoStop [0]{.\EOS\space}%
\providecommand \EOS [0]{\spacefactor3000\relax}%
\providecommand \BibitemShut  [1]{\csname bibitem#1\endcsname}%
\let\auto@bib@innerbib\@empty
\bibitem [{\citenamefont {Nobre}\ \emph {et~al.}(2011)\citenamefont {Nobre},
  \citenamefont {Rego-Monteiro},\ and\ \citenamefont {Tsallis}}]{tsal}%
  \BibitemOpen
  \bibfield  {author} {\bibinfo {author} {\bibfnamefont {F.~D.}\ \bibnamefont
  {Nobre}}, \bibinfo {author} {\bibfnamefont {M.~A.}\ \bibnamefont
  {Rego-Monteiro}}, \ and\ \bibinfo {author} {\bibfnamefont {C.}~\bibnamefont
  {Tsallis}},\ }\href@noop {} {\bibfield  {journal} {\bibinfo  {journal} {Phys.
  Rev. Lett.}\ }\textbf {\bibinfo {volume} {106}},\ \bibinfo {pages} {140601}
  (\bibinfo {year} {2011})}\BibitemShut {NoStop}%
\bibitem [{\citenamefont {Filho}\ \emph {et~al.}(2011)\citenamefont {Filho},
  \citenamefont {Almeida},\ and\ \citenamefont {de~Farias}}]{rcf}%
  \BibitemOpen
  \bibfield  {author} {\bibinfo {author} {\bibfnamefont {R.~N.~C.}\
  \bibnamefont {Filho}}, \bibinfo {author} {\bibfnamefont {M.~P.}\ \bibnamefont
  {Almeida}}, \ and\ \bibinfo {author} {\bibfnamefont {G.~A.}\ \bibnamefont
  {de~Farias}},\ }\href@noop {} {\bibfield  {journal} {\bibinfo  {journal}
  {Phys. Rev. A}\ }\textbf {\bibinfo {volume} {84}},\ \bibinfo {pages} {050102}
  (\bibinfo {year} {2011})}\BibitemShut {NoStop}%
\bibitem [{\citenamefont {Ginzburg}\ and\ \citenamefont {Landau}(1950)}]{glt}%
  \BibitemOpen
  \bibfield  {author} {\bibinfo {author} {\bibfnamefont {V.~L.}\ \bibnamefont
  {Ginzburg}}\ and\ \bibinfo {author} {\bibfnamefont {L.~D.}\ \bibnamefont
  {Landau}},\ }\href@noop {} {\bibfield  {journal} {\bibinfo  {journal} {Zh.
  Eksp. Teor. Fiz.}\ }\textbf {\bibinfo {volume} {20}},\ \bibinfo {pages}
  {1064} (\bibinfo {year} {1950})}\BibitemShut {NoStop}%
\bibitem [{\citenamefont {Gross}(1961)}]{gpe1}%
  \BibitemOpen
  \bibfield  {author} {\bibinfo {author} {\bibfnamefont {E.~P.}\ \bibnamefont
  {Gross}},\ }\href@noop {} {\bibfield  {journal} {\bibinfo  {journal} {Il
  Nuovo Cimento}\ }\textbf {\bibinfo {volume} {20}},\ \bibinfo {pages} {454}
  (\bibinfo {year} {1961})}\BibitemShut {NoStop}%
\bibitem [{\citenamefont {Pitaevskii}(1961)}]{gpe2}%
  \BibitemOpen
  \bibfield  {author} {\bibinfo {author} {\bibfnamefont {L.~P.}\ \bibnamefont
  {Pitaevskii}},\ }\href@noop {} {\bibfield  {journal} {\bibinfo  {journal}
  {Sov. Phys. JETP}\ }\textbf {\bibinfo {volume} {13}},\ \bibinfo {pages} {451}
  (\bibinfo {year} {1961})}\BibitemShut {NoStop}%
\bibitem [{\citenamefont {Som}\ \emph {et~al.}(1979)\citenamefont {Som},
  \citenamefont {Gupta},\ and\ \citenamefont {Dasgupta}}]{som1979coupled}%
  \BibitemOpen
  \bibfield  {author} {\bibinfo {author} {\bibfnamefont {B.}~\bibnamefont
  {Som}}, \bibinfo {author} {\bibfnamefont {M.}~\bibnamefont {Gupta}}, \ and\
  \bibinfo {author} {\bibfnamefont {B.}~\bibnamefont {Dasgupta}},\ }\href@noop
  {} {\bibfield  {journal} {\bibinfo  {journal} {Phys. Lett. A}\ }\textbf
  {\bibinfo {volume} {72}},\ \bibinfo {pages} {111} (\bibinfo {year}
  {1979})}\BibitemShut {NoStop}%
\bibitem [{\citenamefont {Gedalin}\ \emph {et~al.}(1997)\citenamefont
  {Gedalin}, \citenamefont {Scott},\ and\ \citenamefont
  {Band}}]{gedalin1997optical}%
  \BibitemOpen
  \bibfield  {author} {\bibinfo {author} {\bibfnamefont {M.}~\bibnamefont
  {Gedalin}}, \bibinfo {author} {\bibfnamefont {T.}~\bibnamefont {Scott}}, \
  and\ \bibinfo {author} {\bibfnamefont {Y.}~\bibnamefont {Band}},\ }\href@noop
  {} {\bibfield  {journal} {\bibinfo  {journal} {Phys. Rev. Lett.}\ }\textbf
  {\bibinfo {volume} {78}},\ \bibinfo {pages} {448} (\bibinfo {year}
  {1997})}\BibitemShut {NoStop}%
\bibitem [{\citenamefont {Nore}\ \emph {et~al.}(1993)\citenamefont {Nore},
  \citenamefont {Brachet},\ and\ \citenamefont {Fauve}}]{nore1993numerical}%
  \BibitemOpen
  \bibfield  {author} {\bibinfo {author} {\bibfnamefont {C.}~\bibnamefont
  {Nore}}, \bibinfo {author} {\bibfnamefont {M.}~\bibnamefont {Brachet}}, \
  and\ \bibinfo {author} {\bibfnamefont {S.}~\bibnamefont {Fauve}},\
  }\href@noop {} {\bibfield  {journal} {\bibinfo  {journal} {Physica D:
  Nonlinear Phenomena}\ }\textbf {\bibinfo {volume} {65}},\ \bibinfo {pages}
  {154} (\bibinfo {year} {1993})}\BibitemShut {NoStop}%
\bibitem [{\citenamefont {Baelus}\ and\ \citenamefont
  {Peeters}(2002)}]{PhysRevB.65.104515}%
  \BibitemOpen
  \bibfield  {author} {\bibinfo {author} {\bibfnamefont {B.}~\bibnamefont
  {Baelus}}\ and\ \bibinfo {author} {\bibfnamefont {F.}~\bibnamefont
  {Peeters}},\ }\href {\doibase 10.1103/PhysRevB.65.104515} {\bibfield
  {journal} {\bibinfo  {journal} {Phys. Rev. B}\ }\textbf {\bibinfo {volume}
  {65}},\ \bibinfo {pages} {104515} (\bibinfo {year} {2002})}\BibitemShut
  {NoStop}%
\bibitem [{\citenamefont {Kelley}(1995)}]{kelley1995iterative}%
  \BibitemOpen
  \bibfield  {author} {\bibinfo {author} {\bibfnamefont {C.}~\bibnamefont
  {Kelley}},\ }\href@noop {} {\emph {\bibinfo {title} {Iterative methods for
  linear and nonlinear equations}}}\ (\bibinfo  {publisher} {Society for
  Industrial Mathematics},\ \bibinfo {year} {1995})\BibitemShut {NoStop}%
\bibitem [{\citenamefont {Knoll}\ and\ \citenamefont
  {Keyes}(2004)}]{knoll2004jacobian}%
  \BibitemOpen
  \bibfield  {author} {\bibinfo {author} {\bibfnamefont {D.}~\bibnamefont
  {Knoll}}\ and\ \bibinfo {author} {\bibfnamefont {D.}~\bibnamefont {Keyes}},\
  }\href@noop {} {\bibfield  {journal} {\bibinfo  {journal} {Journal of
  Computational Physics}\ }\textbf {\bibinfo {volume} {193}},\ \bibinfo {pages}
  {357} (\bibinfo {year} {2004})}\BibitemShut {NoStop}%
\bibitem [{\citenamefont {Krauskopf}(2007)}]{Krauskopf2007}%
  \BibitemOpen
  \bibfield  {author} {\bibinfo {author} {\bibfnamefont {B.}~\bibnamefont
  {Krauskopf}},\ }\href@noop {} {\emph {\bibinfo {title} {{Numerical
  Continuation Methods for Dynamical Systems: Path following and boundary value
  problems}}}}\ (\bibinfo  {publisher} {Springer Verlag},\ \bibinfo {year}
  {2007})\BibitemShut {NoStop}%
\bibitem [{\citenamefont {Schweigert}\ and\ \citenamefont
  {Peeters}(1999)}]{schw1999}%
  \BibitemOpen
  \bibfield  {author} {\bibinfo {author} {\bibfnamefont {V.~A.}\ \bibnamefont
  {Schweigert}}\ and\ \bibinfo {author} {\bibfnamefont {F.~M.}\ \bibnamefont
  {Peeters}},\ }\href@noop {} {\bibfield  {journal} {\bibinfo  {journal} {Phys.
  Rev. Lett.}\ }\textbf {\bibinfo {volume} {83}},\ \bibinfo {pages} {2409}
  (\bibinfo {year} {1999})}\BibitemShut {NoStop}%
\bibitem [{\citenamefont {Schl\"omer}\ \emph {et~al.}(2012)\citenamefont
  {Schl\"omer}, \citenamefont {Avitabile},\ and\ \citenamefont
  {Vanroose}}]{SAV:2012:NBS}%
  \BibitemOpen
  \bibfield  {author} {\bibinfo {author} {\bibfnamefont {N.}~\bibnamefont
  {Schl\"omer}}, \bibinfo {author} {\bibfnamefont {D.}~\bibnamefont
  {Avitabile}}, \ and\ \bibinfo {author} {\bibfnamefont {W.}~\bibnamefont
  {Vanroose}},\ }\href@noop {} {\bibfield  {journal} {\bibinfo  {journal} {SIAM
  Journal on Applied Dynamical Systems}\ }\textbf {\bibinfo {volume} {11}},\
  \bibinfo {pages} {447} (\bibinfo {year} {2012})}\BibitemShut {NoStop}%
\bibitem [{\citenamefont {Taha}\ and\ \citenamefont
  {Ablowitz}(1984)}]{Taha1984203}%
  \BibitemOpen
  \bibfield  {author} {\bibinfo {author} {\bibfnamefont {T.~R.}\ \bibnamefont
  {Taha}}\ and\ \bibinfo {author} {\bibfnamefont {M.~I.}\ \bibnamefont
  {Ablowitz}},\ }\href@noop {} {\bibfield  {journal} {\bibinfo  {journal}
  {Journal of Computational Physics}\ }\textbf {\bibinfo {volume} {55}},\
  \bibinfo {pages} {203} (\bibinfo {year} {1984})}\BibitemShut {NoStop}%
\bibitem [{\citenamefont {Sanz-Serna}(1984)}]{sanz1984methods}%
  \BibitemOpen
  \bibfield  {author} {\bibinfo {author} {\bibfnamefont {J.}~\bibnamefont
  {Sanz-Serna}},\ }\href@noop {} {\bibfield  {journal} {\bibinfo  {journal}
  {Mathematics of Computation}\ }\textbf {\bibinfo {volume} {43}},\ \bibinfo
  {pages} {21} (\bibinfo {year} {1984})}\BibitemShut {NoStop}%
\bibitem [{\citenamefont {Chang}\ \emph {et~al.}(1999)\citenamefont {Chang},
  \citenamefont {Jia},\ and\ \citenamefont {Sun}}]{Chang1999397}%
  \BibitemOpen
  \bibfield  {author} {\bibinfo {author} {\bibfnamefont {Q.}~\bibnamefont
  {Chang}}, \bibinfo {author} {\bibfnamefont {E.}~\bibnamefont {Jia}}, \ and\
  \bibinfo {author} {\bibfnamefont {W.}~\bibnamefont {Sun}},\ }\href@noop {}
  {\bibfield  {journal} {\bibinfo  {journal} {Journal of Computational
  Physics}\ }\textbf {\bibinfo {volume} {148}},\ \bibinfo {pages} {397}
  (\bibinfo {year} {1999})}\BibitemShut {NoStop}%
\bibitem [{\citenamefont {Ruprecht}\ \emph {et~al.}(1995)\citenamefont
  {Ruprecht}, \citenamefont {Holland}, \citenamefont {Burnett},\ and\
  \citenamefont {Edwards}}]{PhysRevA.51.4704}%
  \BibitemOpen
  \bibfield  {author} {\bibinfo {author} {\bibfnamefont {P.~A.}\ \bibnamefont
  {Ruprecht}}, \bibinfo {author} {\bibfnamefont {M.~J.}\ \bibnamefont
  {Holland}}, \bibinfo {author} {\bibfnamefont {K.}~\bibnamefont {Burnett}}, \
  and\ \bibinfo {author} {\bibfnamefont {M.}~\bibnamefont {Edwards}},\
  }\href@noop {} {\bibfield  {journal} {\bibinfo  {journal} {Phys. Rev. A}\
  }\textbf {\bibinfo {volume} {51}},\ \bibinfo {pages} {4704} (\bibinfo {year}
  {1995})}\BibitemShut {NoStop}%
\bibitem [{\citenamefont {Bao}\ \emph {et~al.}(2003)\citenamefont {Bao},
  \citenamefont {Jaksch},\ and\ \citenamefont {Markowich}}]{bao2003numerical}%
  \BibitemOpen
  \bibfield  {author} {\bibinfo {author} {\bibfnamefont {W.}~\bibnamefont
  {Bao}}, \bibinfo {author} {\bibfnamefont {D.}~\bibnamefont {Jaksch}}, \ and\
  \bibinfo {author} {\bibfnamefont {P.}~\bibnamefont {Markowich}},\ }\href@noop
  {} {\bibfield  {journal} {\bibinfo  {journal} {Journal of Computational
  Physics}\ }\textbf {\bibinfo {volume} {187}},\ \bibinfo {pages} {318}
  (\bibinfo {year} {2003})}\BibitemShut {NoStop}%
\bibitem [{\citenamefont {Schweigert}\ and\ \citenamefont
  {Peeters}(1998)}]{schwPRL}%
  \BibitemOpen
  \bibfield  {author} {\bibinfo {author} {\bibfnamefont {V.~A.}\ \bibnamefont
  {Schweigert}}\ and\ \bibinfo {author} {\bibfnamefont {F.~M.}\ \bibnamefont
  {Peeters}},\ }\href@noop {} {\bibfield  {journal} {\bibinfo  {journal} {Phys.
  Rev. Lett.}\ }\textbf {\bibinfo {volume} {81}},\ \bibinfo {pages} {2783}
  (\bibinfo {year} {1998})}\BibitemShut {NoStop}%
\bibitem [{\citenamefont {Deo}\ \emph {et~al.}(1997)\citenamefont {Deo},
  \citenamefont {Schweigert}, \citenamefont {Peeters},\ and\ \citenamefont
  {Geim}}]{PhysRevLett.79.4653}%
  \BibitemOpen
  \bibfield  {author} {\bibinfo {author} {\bibfnamefont {P.~S.}\ \bibnamefont
  {Deo}}, \bibinfo {author} {\bibfnamefont {V.}~\bibnamefont {Schweigert}},
  \bibinfo {author} {\bibfnamefont {F.}~\bibnamefont {Peeters}}, \ and\
  \bibinfo {author} {\bibfnamefont {A.}~\bibnamefont {Geim}},\ }\href {\doibase
  10.1103/PhysRevLett.79.4653} {\bibfield  {journal} {\bibinfo  {journal}
  {Phys. Rev. Lett.}\ }\textbf {\bibinfo {volume} {79}},\ \bibinfo {pages}
  {4653} (\bibinfo {year} {1997})}\BibitemShut {NoStop}%
\bibitem [{\citenamefont {Milo\v{s}evi\'{c}}\ and\ \citenamefont
  {Geurts}(2010)}]{geurts}%
  \BibitemOpen
  \bibfield  {author} {\bibinfo {author} {\bibfnamefont {M.~V.}\ \bibnamefont
  {Milo\v{s}evi\'{c}}}\ and\ \bibinfo {author} {\bibfnamefont {R.}~\bibnamefont
  {Geurts}},\ }\href@noop {} {\bibfield  {journal} {\bibinfo  {journal}
  {Physica C}\ }\textbf {\bibinfo {volume} {470}},\ \bibinfo {pages} {791}
  (\bibinfo {year} {2010})}\BibitemShut {NoStop}%
\bibitem [{\citenamefont {Saad}(2003)}]{saad2003iterative}%
  \BibitemOpen
  \bibfield  {author} {\bibinfo {author} {\bibfnamefont {Y.}~\bibnamefont
  {Saad}},\ }\href@noop {} {\emph {\bibinfo {title} {Iterative methods for
  sparse linear systems}}}\ (\bibinfo  {publisher} {Society for Industrial
  Mathematics},\ \bibinfo {year} {2003})\BibitemShut {NoStop}%
\bibitem [{\citenamefont {Greenbaum}(1997)}]{greenbaum1997iterative}%
  \BibitemOpen
  \bibfield  {author} {\bibinfo {author} {\bibfnamefont {A.}~\bibnamefont
  {Greenbaum}},\ }\href@noop {} {\emph {\bibinfo {title} {Iterative methods for
  solving linear systems}}},\ Vol.~\bibinfo {volume} {17}\ (\bibinfo
  {publisher} {Society for Industrial Mathematics},\ \bibinfo {year}
  {1997})\BibitemShut {NoStop}%
\bibitem [{\citenamefont {Schl\"omer}\ and\ \citenamefont
  {Vanroose}(2012)}]{SV:2012:OLS}%
  \BibitemOpen
  \bibfield  {author} {\bibinfo {author} {\bibfnamefont {N.}~\bibnamefont
  {Schl\"omer}}\ and\ \bibinfo {author} {\bibfnamefont {W.}~\bibnamefont
  {Vanroose}},\ }\href@noop {} {\bibfield  {journal} {\bibinfo  {journal}
  {Journal of Computational Physics (submitted)}\ } (\bibinfo {year}
  {2012})}\BibitemShut {NoStop}%
\bibitem [{\citenamefont {Trottenberg}\ \emph {et~al.}(2001)\citenamefont
  {Trottenberg}, \citenamefont {Oosterlee},\ and\ \citenamefont
  {Sch\"uller}}]{trottenberg2001multigrid}%
  \BibitemOpen
  \bibfield  {author} {\bibinfo {author} {\bibfnamefont {U.}~\bibnamefont
  {Trottenberg}}, \bibinfo {author} {\bibfnamefont {C.}~\bibnamefont
  {Oosterlee}}, \ and\ \bibinfo {author} {\bibfnamefont {A.}~\bibnamefont
  {Sch\"uller}},\ }\href@noop {} {\emph {\bibinfo {title} {Multigrid}}}\
  (\bibinfo  {publisher} {Academic Press},\ \bibinfo {year} {2001})\BibitemShut
  {NoStop}%
\bibitem [{\citenamefont {Keller}\ \emph {et~al.}(1987)\citenamefont {Keller},
  \citenamefont {Nandakumaran},\ and\ \citenamefont
  {Ramaswamy}}]{keller1987lectures}%
  \BibitemOpen
  \bibfield  {author} {\bibinfo {author} {\bibfnamefont {H.}~\bibnamefont
  {Keller}}, \bibinfo {author} {\bibfnamefont {A.}~\bibnamefont
  {Nandakumaran}}, \ and\ \bibinfo {author} {\bibfnamefont {M.}~\bibnamefont
  {Ramaswamy}},\ }\href@noop {} {\bibfield  {journal} {\bibinfo  {journal}
  {Applied Mathematics}\ }\textbf {\bibinfo {volume} {217}},\ \bibinfo {pages}
  {50} (\bibinfo {year} {1987})}\BibitemShut {NoStop}%
\bibitem [{\citenamefont {Du}\ \emph {et~al.}(1992)\citenamefont {Du},
  \citenamefont {Gunzburger},\ and\ \citenamefont {Peterson}}]{DGP:1992:AAG}%
  \BibitemOpen
  \bibfield  {author} {\bibinfo {author} {\bibfnamefont {Q.}~\bibnamefont
  {Du}}, \bibinfo {author} {\bibfnamefont {M.~D.}\ \bibnamefont {Gunzburger}},
  \ and\ \bibinfo {author} {\bibfnamefont {J.~S.}\ \bibnamefont {Peterson}},\
  }\href@noop {} {\bibfield  {journal} {\bibinfo  {journal} {SIAM Rev.}\
  }\textbf {\bibinfo {volume} {34}},\ \bibinfo {pages} {54} (\bibinfo {year}
  {1992})}\BibitemShut {NoStop}%
\bibitem [{\citenamefont {Bulgac}\ \emph {et~al.}(2011)\citenamefont {Bulgac},
  \citenamefont {Luo}, \citenamefont {Magierski}, \citenamefont {Roche},\ and\
  \citenamefont {Yu}}]{ScVort}%
  \BibitemOpen
  \bibfield  {author} {\bibinfo {author} {\bibfnamefont {A.}~\bibnamefont
  {Bulgac}}, \bibinfo {author} {\bibfnamefont {Y.-L.}\ \bibnamefont {Luo}},
  \bibinfo {author} {\bibfnamefont {P.}~\bibnamefont {Magierski}}, \bibinfo
  {author} {\bibfnamefont {K.~J.}\ \bibnamefont {Roche}}, \ and\ \bibinfo
  {author} {\bibfnamefont {Y.}~\bibnamefont {Yu}},\ }\href@noop {} {\bibfield
  {journal} {\bibinfo  {journal} {Science}\ }\textbf {\bibinfo {volume}
  {332}},\ \bibinfo {pages} {1288} (\bibinfo {year} {2011})}\BibitemShut
  {NoStop}%
\bibitem [{\citenamefont {Schl\"omer}(2012)}]{schloemer:figshare}%
  \BibitemOpen
  \bibfield  {author} {\bibinfo {author} {\bibfnamefont {N.}~\bibnamefont
  {Schl\"omer}},\ }\href {http://dx.doi.org/10.6084/m9.figshare.96108}
  {\enquote {\bibinfo {title} {Vortex patterns for a cubic extreme-type-{II}
  superconductor with spherical cavity},}\ } (\bibinfo {year}
  {2012})\BibitemShut {NoStop}%
\bibitem [{\citenamefont {Heroux}\ \emph {et~al.}(2005)\citenamefont {Heroux},
  \citenamefont {Bartlett}, \citenamefont {Howle}, \citenamefont {Hoekstra},
  \citenamefont {Hu}, \citenamefont {Kolda}, \citenamefont {Lehoucq},
  \citenamefont {Long}, \citenamefont {Pawlowski}, \citenamefont {Phipps},
  \citenamefont {Salinger}, \citenamefont {Thornquist}, \citenamefont
  {Tuminaro}, \citenamefont {Willenbring}, \citenamefont {Williams},\ and\
  \citenamefont {Stanley}}]{1089021}%
  \BibitemOpen
  \bibfield  {author} {\bibinfo {author} {\bibfnamefont {M.~A.}\ \bibnamefont
  {Heroux}}, \bibinfo {author} {\bibfnamefont {R.~A.}\ \bibnamefont
  {Bartlett}}, \bibinfo {author} {\bibfnamefont {V.~E.}\ \bibnamefont {Howle}},
  \bibinfo {author} {\bibfnamefont {R.~J.}\ \bibnamefont {Hoekstra}}, \bibinfo
  {author} {\bibfnamefont {J.~J.}\ \bibnamefont {Hu}}, \bibinfo {author}
  {\bibfnamefont {T.~G.}\ \bibnamefont {Kolda}}, \bibinfo {author}
  {\bibfnamefont {R.~B.}\ \bibnamefont {Lehoucq}}, \bibinfo {author}
  {\bibfnamefont {K.~R.}\ \bibnamefont {Long}}, \bibinfo {author}
  {\bibfnamefont {R.~P.}\ \bibnamefont {Pawlowski}}, \bibinfo {author}
  {\bibfnamefont {E.~T.}\ \bibnamefont {Phipps}}, \bibinfo {author}
  {\bibfnamefont {A.~G.}\ \bibnamefont {Salinger}}, \bibinfo {author}
  {\bibfnamefont {H.~K.}\ \bibnamefont {Thornquist}}, \bibinfo {author}
  {\bibfnamefont {R.~S.}\ \bibnamefont {Tuminaro}}, \bibinfo {author}
  {\bibfnamefont {J.~M.}\ \bibnamefont {Willenbring}}, \bibinfo {author}
  {\bibfnamefont {A.}~\bibnamefont {Williams}}, \ and\ \bibinfo {author}
  {\bibfnamefont {K.~S.}\ \bibnamefont {Stanley}},\ }\href@noop {} {\bibfield
  {journal} {\bibinfo  {journal} {ACM Trans. Math. Softw.}\ }\textbf {\bibinfo
  {volume} {31}},\ \bibinfo {pages} {397} (\bibinfo {year} {2005})}\BibitemShut
  {NoStop}%
\bibitem [{\citenamefont {Doria}\ \emph {et~al.}(2007)\citenamefont {Doria},
  \citenamefont {Romaguera}, \citenamefont {Milo\v{s}evi\'{c}},\ and\
  \citenamefont {Peeters}}]{mauro}%
  \BibitemOpen
  \bibfield  {author} {\bibinfo {author} {\bibfnamefont {M.~M.}\ \bibnamefont
  {Doria}}, \bibinfo {author} {\bibfnamefont {A.~R.~C.}\ \bibnamefont
  {Romaguera}}, \bibinfo {author} {\bibfnamefont {M.~V.}\ \bibnamefont
  {Milo\v{s}evi\'{c}}}, \ and\ \bibinfo {author} {\bibfnamefont {F.~M.}\
  \bibnamefont {Peeters}},\ }\href@noop {} {\bibfield  {journal} {\bibinfo
  {journal} {Europhys. Lett.}\ }\textbf {\bibinfo {volume} {79}},\ \bibinfo
  {pages} {47006} (\bibinfo {year} {2007})}\BibitemShut {NoStop}%
\bibitem [{\citenamefont {Faddeev}\ and\ \citenamefont {Niemi}(1997)}]{knot1}%
  \BibitemOpen
  \bibfield  {author} {\bibinfo {author} {\bibfnamefont {L.~D.}\ \bibnamefont
  {Faddeev}}\ and\ \bibinfo {author} {\bibfnamefont {A.~J.}\ \bibnamefont
  {Niemi}},\ }\href@noop {} {\bibfield  {journal} {\bibinfo  {journal} {Nature
  (London)}\ }\textbf {\bibinfo {volume} {387}},\ \bibinfo {pages} {58}
  (\bibinfo {year} {1997})}\BibitemShut {NoStop}%
\bibitem [{\citenamefont {Moore}\ \emph {et~al.}(2008)\citenamefont {Moore},
  \citenamefont {Ran},\ and\ \citenamefont {Wen}}]{knot2}%
  \BibitemOpen
  \bibfield  {author} {\bibinfo {author} {\bibfnamefont {J.~E.}\ \bibnamefont
  {Moore}}, \bibinfo {author} {\bibfnamefont {Y.}~\bibnamefont {Ran}}, \ and\
  \bibinfo {author} {\bibfnamefont {X.-G.}\ \bibnamefont {Wen}},\ }\href@noop
  {} {\bibfield  {journal} {\bibinfo  {journal} {Phys. Rev. Lett.}\ }\textbf
  {\bibinfo {volume} {101}},\ \bibinfo {pages} {186805} (\bibinfo {year}
  {2008})}\BibitemShut {NoStop}%
\bibitem [{\citenamefont {Scott}(2007)}]{nonlin}%
  \BibitemOpen
  \bibfield  {author} {\bibinfo {author} {\bibfnamefont {A.~C.}\ \bibnamefont
  {Scott}},\ }\href@noop {} {\emph {\bibinfo {title} {The Nonlinear
  Universe}}}\ (\bibinfo  {publisher} {Springer, Berlin},\ \bibinfo {year}
  {2007})\BibitemShut {NoStop}%
\end{thebibliography}%
\end{document}